\documentclass[twocolumn,showpacs,preprintnumbers,amssymb,aps,pra,]{revtex4}

\usepackage{amsmath}
\usepackage{graphicx}

\begin{document}

\title{Ground State Properties of a Tonks-Girardeau Gas in a Split Trap}

\author{J.~Goold}
\email{jgoold@phys.ucc.ie}
\author{Th.~Busch}
\affiliation{ Department of Physics,
  National University of Ireland, UCC, Cork, Republic of Ireland}%

\date{\today}

\begin{abstract}
  We determine the exact many-body properties of a bosonic
  Tonks-Girardeau gas confined in a harmonic potential with a tunable
  $\delta$-function barrier at the trap center. This is done by
  calculating the reduced single particle density matrix, the
  pair-distribution function, and momentum distribution of the gas as
  a function of barrier strength and particle number. With increasing
  barrier height we find that the ground state occupation in a
  diagonal basis diverges from the $\sqrt N$ behavior that is expected
  for the case of a simple harmonic trap. In fact, the scaling of the
  occupation number depends on whether one has an even or odd number
  of particles. Since this quantity is a measure of the
  coherence of our sample we show how the odd-even effect manifests
  itself in both the momentum distribution of the Bose gas and
  interference fringe visibility during free temporal evolution.
\end{abstract}


%
%

\maketitle

\section{Introduction}

The last two decades have seen considerable experimental and
theoretical activity and progress in the area of cooling and trapping
of neutral atoms. Several crowning achievements have been produced,
including the realization of Bose-Einstein condensation
\cite{Dalfovo:99}, the creation of periodic arrays of single atoms
\cite{Bloch:07} and the observation of superfluid phases in ultra-cold
Fermi gases \cite{Giorgini:07}. While research in this area is
interesting from a fundamental point of view, ultracold atoms are also
well-suited candidates to observe concepts and ideas in quantum
information \cite{Monroe:02}. This is due to the fact that cold atomic
samples are often well isolated from the environment, while being
highly controllable at the same time, which is of paramount importance
if one wants to create and work with fragile many-body states.

Two experimental advances have recently opened up the possibility to
create and carry out experiments in strongly correlated quantum gases.
The first one is due to optical lattices and atom chip traps, which
can produce tightly confined potentials in selective directions of
space. This allows one to limit the atom's degrees of freedom and
thereby create effectively lower-dimensional systems
\cite{Paredes:04,Kinoshita:04}. Secondly, by the using Feshbach
resonances or by tuning of the effective mass of particles moving in a
periodic potential \cite{Paredes:04}, the inter-particle scattering
length can be tuned to almost every value desired.

Combining these techniques has permitted the experimental realization
of atomic gases in the so-called Tonks-Girardeau (TG) regime
\cite{Paredes:04, Kinoshita:04}. A TG gas is defined to be a
one-dimensional, strongly correlated gas consisting of bosons that
interact via a hard-core potential
\cite{Tonks:36,Girardeau:60,Yukalov:05}. In the limit of point-like
inter-particle interactions Girardeau found that such a model can be
solved exactly by mapping it to an ideal, spinless fermionic system
and he was the first to point out that a gas of strongly-interacting
bosons can thereby acquire certain fermionic properties
\cite{Girardeau:60}. More recently it was found that the above case is
just a special case of a general mapping theorem between bosons and
fermions in one-dimension, which can interact with finite strength
\cite{Granger:04}.

In order to find the many-particle solutions of a given geometry for
the TG gas (or for non-interacting fermions) one must first know the
exact single particle eigenstates. However, since the list of exactly
solvable single particle problems in quantum mechanics is limited,
there is also only a small number of many-particle problems in the
Tonks limit that can be exactly solved. Recently an exact solution for
the experimentally important harmonic potential was found
\cite{Wright:00,Girardeau:01} and here we will add another example to
the list by describing a double well setting. As we aim for exact
solvability, we have chosen the model of the $\delta$-split trap
\cite{Busch:03}, which comprises of a gas trapped in a harmonic
oscillator potential split in the center by a point-like repulsive
potential. In a recent paper the case of a boson pair was rigorously
analysed in such a split trap for a range of interaction strengths up
to and including the Tonks limit \cite{Murphy:07}. It has also been suggested in \cite{Segev:06} that
the $\delta$-split trap could be used to excite dark soliton like structures
in a Tonks-Girardeau gas.

Investigating the model of the $\delta$-split potential can be
justified in several ways. Firstly it can be seen as an idealized
model of a realistic double well situation where the height of the
barrier is related to the area of a physical potential. Comparison of
our results with recent numerical simulations show that the
qualitative behavior we find here persists for reasonable and
realistic values of finite sized splitting potentials
\cite{Murphy:08,Zollner:06,Zollner2:06}. Alternatively, a point like potential can
be a good approximation to describe a strongly localized impurity
within the bosonic gas \cite{Castin:05,Rojo:06}.

Our main findings concern the coherence inherent in such samples. For
the harmonically trapped case and in the thermodynamic limit it is
known that the groundstate occupation scales as $\sqrt{N}$
\cite{Forrester:03}. However, for smaller samples it has been
demonstrated that deviations from the $\sqrt{N}$ behavior exist and
that macroscopic coherence effects can still be present
\cite{Girardeau:01}. Therefore in this work we put particular emphasis
on examining the ground-state occupation fraction as we increase the
height of the central $\delta$-barrier.

The paper is organized as follows. In
Sec.~\ref{sect:model_hamiltonian} we introduce the many-body
Hamiltonian, describe the associated single particle eigenfunctions
and eigenvalues and review the Fermi-Bose mapping theorem.
Sec.~\ref{sect:manybody} explores the physical many-body properties of
our model by calculating the reduced single particle density matrix
(RSPDM) and the pair-correlation function. In Sec.~\ref{sect:occu} we
investigate the influence of the splitting strength on the ground
state occupation number and relate our results for this behavior to
two experimentally realizable quantities, namely the momentum
distribution and interference patterns. Finally, in
Sec.~\ref{sect:conclusions} we make concluding remarks.

\section{Model Hamiltonian and the Fermi-Bose Mapping Theorem}
\label{sect:model_hamiltonian}

\subsection{System Hamiltonian}
We consider a gas of $N$ bosons trapped in a tight atomic
waveguide. The waveguide restricts the dynamic of the gas strongly in
the transversal directions, such that in the low-temperature limit we
can restrict our model to the longitudinal direction only
\cite{Olshanii:98}. In this direction we then consider a
$\delta$-split harmonic potential such that at low linear density the
many-particle Hamiltonian can be written as
\begin{align}
  \label{eq:Hamiltonian}
  \mathcal{H}=&\sum_{n=1}^N
         \left(-\frac{\hbar^2}{2m}\frac{\partial^2}{\partial x_n^2}
               +\frac{1}{2}m\omega^2x_n^2+\kappa\delta(x_n)\right) \nonumber\\
              &\quad+\sum_{i<j}V(|x_i-x_j|)\;.
\end{align}
Here $m$ is the mass of a single atom, $\omega$ the frequency of the
harmonic potential and $\kappa$ is the strength of the point-like
splitting potential, which is located at $x=0$.  Since we assume low
densities, only elastic two-particle collision have to be considered
and we can restrict the interaction potential to depend only on the
relative distances. For bosonic systems at low temperatures the atomic
interaction potential itself can be well approximated by a point-like
potential
\begin{equation}
 \label{eq:interaction}
 V(|x_i-x_j|)=g_{1D}\delta(|x_i-x_j|)\;,
\end{equation}
where $g_{1D}$ is the 1D coupling constant. This approximation is well
justified for non-resonant situations and the only reminiscence of the
exact potential is given by the three-dimensional s-wave scattering
length, $a_{3D}$. For positive values of $a_{3D}$ the interaction is
repulsive and for negative values of $a_{3D}$ it is
attractive. Finally, the scattering length is related to the
one-dimensional coupling constant via
\begin{equation}
  g_{1D}=\frac{4\hbar^2 a_{3D}}{ma_\perp}
         \left(a_\perp-Ca_{3D}\right)^{-1}\;,
\end{equation}
where $C$ is a constant of value $C=1.4603\dots$ \cite{Olshanii:98}.

\subsection{Eigenstates and eigenvalues of the $\delta$-split trap}

The single particle eigenstates of the delta-split harmonic oscillator
have recently been discussed in detail \cite{Busch:98} and we will
briefly review them here for completeness.  To do this we rescale the single particle part of the Hamiltonian
\eqref{eq:Hamiltonian}
\begin{equation}
 \label{eq:sin_part_ham_scaled}
  h = -\frac{1}{2}\frac{\partial^2}{\partial \bar{x}^2} 
      + \frac{1}{2} \bar{x}^2 + \bar{\kappa} \delta (\bar{x})\;, 
\end{equation}
where all length are in units of the ground state size,
$a_0=\sqrt{\hbar/m\omega}$, and all energies in terms of the
oscillator energy, $\hbar\omega$. This leads to a new scaled length given by $x=\bar{x}a_0$ and scaled barrier strength
given by $\bar{\kappa} = ( \hbar \omega a_0)^{-1} \kappa$. For notational simplicity we shall drop the overbars on all scaled quantities and acknolwege that we are, henceforth, dealing in the scaled units just described. All units used in figure plots in this paper are also in terms of these scaled units. The
time-independent Schr\"odinger equation for this system now reads
\begin{equation}
 \label{eq:tise_scaled}
 h\psi_n(x) = E_n \psi_n(x)\;. 
\end{equation}
It is immediately clear that the odd eigenfunctions of the simple
harmonic oscillator are still good eigenfunctions for the
$\delta$-split oscillator, as they vanish at the exact position of the
disturbance
\begin{equation}
  \label{eq:sin_part_anti_vec}
 \psi_n(x) = \mathcal{N}_n H_n (x) e^{-\frac{x^2}{2}} \;,
             \quad n =1,3,5\ldots\;.
\end{equation}
Here the $H_n(x)$ are the $n^{th}$ order Hermite polynomials and the
$\mathcal{N}_n$ are the associated normalization constants.  The
corresponding energies are given by the eigenvalues of the odd parity
states of the harmonic oscillator, $E_n = \left( n + \frac{1}{2}
\right)$.

The even eigenstates of the simple harmonic oscillator on the other
hand have an extremum at $x=0$. They are therefore strongly influenced
by the splitting potential and can be found to be 
\cite{Busch:98}
\begin{equation}
 \label{eq:sin_part_sym_vec}
 \psi_n (x) = \mathcal{N}_n\; e^{-\frac{x^2}{2}} U \left(
 \frac{1}{4} - \frac{E_n}{2}, \frac{1}{2}, x^2 \right)\;,\quad n =
 0, 2, 4 \ldots\;,
\end{equation}
where the $U(a,b,z)$ are the Kummer functions \cite{Abramowitz:72}. The
corresponding eigenenergies, $E_n$, are determined by the roots of the
implicit relation,
\begin{equation}
 \label{eq:sin_part_sym_val}
 -\kappa = 2 \frac{\Gamma\left( -\frac{E_n}{2} + \frac{3}{4}\right)}
                  {\Gamma\left( -\frac{E_n}{2} + \frac{1}{4}\right)}\;.
\end{equation}
Increasing the barrier height leads to an increase in the energy of
the even eigenstates and in the limit of $\kappa=\infty$ each even
eigenstate becomes energetically degenerate with the next higher lying
odd eigenstate.

Since we know the single particle eigenstates, we can build and solve
the Slater determinant for a system of non-interacting
fermions. Using the Fermi-Bose mapping theorem we can then calculate
the many-particle bosonic wave-function from the fermionic result.

\subsection{The Fermi-Bose Mapping Theorem}
While the original Fermi-Bose mapping theorem only related
strongly interacting bosons to ideal fermions \cite{Girardeau:60}, it
was recently found that the mapping idea can be applied to other
systems as well \cite{Granger:04}. Here we concentrate on the
situation relevant to our system, i.e.~the Tonks limit of infinite,
point-like repulsion $(g_{1D} \rightarrow \infty)$ between bosons.
The main idea is that one can treat the interaction term in
eq.~(\ref{eq:Hamiltonian}) by replacing it with the following boundary
condition on the allowed bosonic wave-function
\begin{equation}
  \label{eq:constraint}
  \Psi_B=0\quad \text{if} \quad |x_i-x_j|=0,\qquad 
\end{equation}
for $i\neq j$ and $1\leq i\leq\ j\leq N$. As this is formally
equivalent the the Pauli exclusion principle, one can solve for the
associated ideal fermionic wave-function
\begin{equation}
  \Psi_F(x_1,\dots,x_N)=
  \frac{1}{\sqrt{N!}} \det_{n,j=1}^N [\psi_n(x_j)], 
\label{psiF}
\end{equation} 
and calculate the bosonic
solution from this by appropriate symmetrization
\begin{equation}
  \Psi_B =  A(x_1,\dots,x_N) \Psi_F(x_1,x_2,\dots,x_N),
\label{mapFB}
\end{equation}
where the unit antisymmetric function is given by \cite{Girardeau:60}
\begin{equation}
A=\prod_{1\leq i < j\leq N} \mbox{sgn}(x_i-x_j)\;.
\label{unitA}
\end{equation}
As we are only interested in the ground state, this last step
simplifies to
\begin{equation}
  \label{eq:wavefunc}
  \Psi_B(x_1,.....x_N)=|\Psi_F(x_1,.....x_N)|\;.
\end{equation}

\section{Many-body properties}
\label{sect:manybody}

With the armory of the mapping theorem and the single particle states
available, we are now in a position to calculate various ground state
properties of the many-particle state as a function of increasing particle
number as well as varying the strength of the central $\delta$ barrier.

\subsection{Reduced single particle density matrices}
\label{subsect:RSPDM}

As laid out in the previous section, the Fermi-Bose mapping theorem
allows for the calculation of the exact many-particle
wave-function. For the case of an infinitely strong barrier this was
recently done analytically in \cite{Busch:03}
\begin{equation}
 \label{eq:PsiB}
 \Psi_B=
  \frac{\tilde C}{\sqrt{N!}} 2^{\frac{N^2}{8}}
  \left[\prod_j^{N/2} e^{-\frac{x_j^2}{4}}|x_j|\right]
  \prod_{(j,k)=(1,j+1)}^{(N/2,N/2)}|x_j^2-x_k^2|\;,
\end{equation}
where $\tilde C$ is the normalisation constant. Although this function
fully characterizes the state of the system, other quantities can be
more useful for obtaining characteristic properties of many-particle
systems. We therefore proceed to calculate the reduced single particle
density matrix (RSPDM), from which expectation values of many
important one-body physical observables such as the momentum
distribution or the von Neumann entropy can easily be obtained.

For the bosonic gas, the RSPDM is defined as
\begin{align}
  \label{eq:rspdm_def}
  \rho(x,x')=\int_{-\infty}^{+\infty} 
             &\Psi_B^\ast(x,x_2,\dots,x_N)\times\nonumber\\
   &\Psi_B(x',x_2,\dots,x_N)\;dx_2\dots dx_N\;,
\end{align}
and we choose its normalisation to be given by $\int \rho(x,x) dx=N$.
While for the simple harmonic oscillator this integral was solved
analytically by Lapeyre {\sl et al.} \cite{Lapeyre:02}, we have to
resort to a numerical evaluation for finite values of $\kappa$. Here
we use an algorithm recently presented by Pezer and Buljan
\cite{Pezer:07} that allows for effective calculation of
eq.~(\ref{eq:rspdm_def}) for large numbers of particles. In fact, our
particle number is only limited by the numerical instabilities when
calculating higher order Kummer functions.

\begin{figure}[t]
   \includegraphics[width=\linewidth]{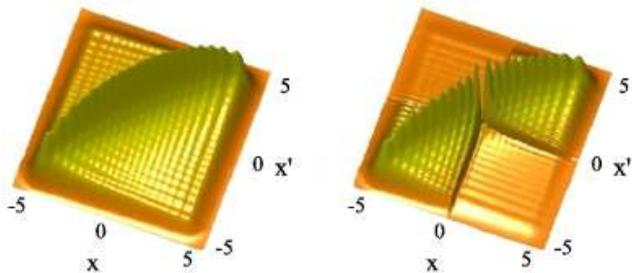}
   \caption{\label{fig:rspdm} (Color online) Reduced single particle density matrices
     $\rho(x,x')$ for a Tonks gas of twenty particles for splitting
     strengths $\kappa=0$ (left) and $\kappa=20$ (right).}
\end{figure}

The RSPDM expresses self correlation and one can view $\rho (x,x')$ as
the probability that, having detected the particle at position $x$, a
second measurement, immediately following the first, will find the
particle at the point $x'$. Classically,
$\rho(x,x')=\delta(x-x')$. The RSPDM for a twenty particle Tonks gas
is shown in Fig.~\ref{fig:rspdm}. In the unsplit case ($\kappa=0$) we
see that most of the density is concentrated along the
diagonal. Increasing $\kappa$ introduces a gap around the position of
the splitting potential $x=x'=0$, along with a reduction of density in
the off diagonal regions. This is due to the suppression of tunnelling
from one side of the barrier to the other as the systems eigenstates
become doubly degenerate in the $\kappa\rightarrow\infty$ limit.

\begin{figure}[t]
   \includegraphics[width=\linewidth]{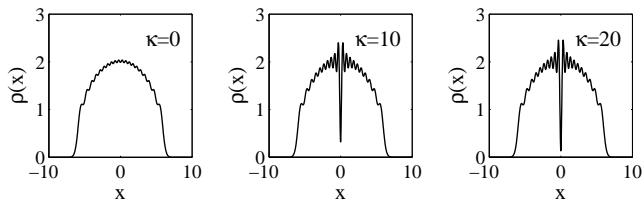}
   \caption{\label{fig:spd} (Color online) Single particle density for a Tonks gas of
     twenty particles for increasing barrier height.}
\end{figure}

The single particle density can be calculated from the RSPDM by taking
$x=x'$
\begin{equation}
  \label{eq:density}
  \rho(x)=N \int_{-\infty}^{+\infty}|\Psi_B(x,x_2\dots,x_N)|^2dx_2\dots dx_N\;,
\end{equation}
and is therefore simply given by the diagonal of $\rho(x,x')$. In
Fig. \ref{fig:spd} we show this quantity for several different values
if $\kappa$. In contrast to the self correlations, one can see that
the splitting effects the density only very locally. The results found
in this case completely agree with the results presented in
\cite{Busch:03}, where they were calculated in a different way by
summing up the single particle densities and symmetrising.

\begin{figure}[t]
  \includegraphics[width=\linewidth]{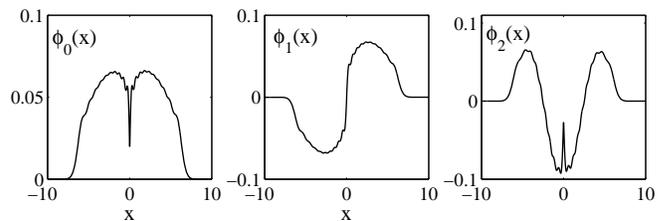}
  \caption{\label{fig:nos} (Color online) The first three natural orbitals for a
    twenty particle Tonks gas in a $\delta$-split trap with splitting
    strength $\kappa=20$.}
\end{figure}

To gain further understanding we next calculate the eigenvalues and
eigenfunctions of the RSPDMs given by
\begin{equation}
  \label{eq:norbs}
  \int_{-\infty}^{\infty}\rho(x,x') \phi_j(x') dx'=\lambda_j \phi_j(x)\;.
\end{equation}
The eigenfunctions $\phi_j(x)$ are known in theoretical chemistry as
\textsl{natural orbitals} and their associated eigenvalues,
$\lambda_j$, represent the occupation number of each \textit{orbital}.
 The first three lowest energy natural orbitals for a twenty particle gas
with a splitting potential of height $\kappa=20$ are displayed in
Fig.~\ref{fig:nos}. One can see the point-like disturbance introduced
by the $\delta$-function into the symmetric orbitals $\phi_0$ and
$\phi_2$, while the anti-symmetric $\phi_1$ is unchanged from the
$\kappa=0$ case (natural orbitals for the $\kappa=0$ case are
displayed in \cite{Girardeau:01}).

These states and their occupation numbers will be used in section
\ref{sect:occu} to calculate the reciprocal momentum distributions and
the ground state occupations.

\subsection{Pair distribution functions}
\label{subsect:spdpdf}

\begin{figure}[t]
   \includegraphics[width=\linewidth]{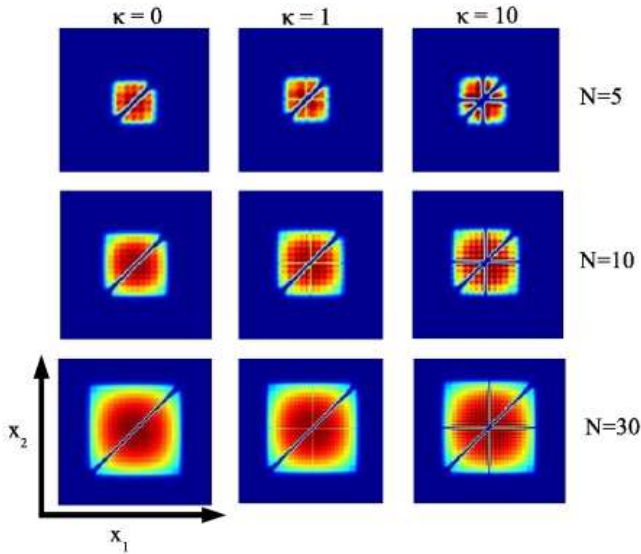}
   \caption{\label{fig:pdf} (Color online) Pair distribution functions $D(x_1,x_2)$
     for a Tonks gas of $N=5,10$ and $30$ particles in
     $\delta$-split trap for splitting strengths $\kappa=0, 1$ and
     $10$. In each plot the horizontal and vertical axes run from $-10$
     to $+10$ in scaled units. }
\end{figure}

The pair-distribution function $D(x_1,x_2)$, is a two-particle
correlation function that describes the probability to measure two
atoms at two given positions at the same time. It is defined in
the following way
\begin{align}
    D(x_1,x_2)&=\mathcal{N}_D\int_{-\infty}^{+\infty} 
                 |\Psi_B(x_1,x_2\dots,x_N)|^{2} dx_3 \dots dx_N\\
              &=\sum_{0\leq n\leq n'\leq N-1}^{N-1} 
                 |\psi_n(x_1) \psi_{n'}(x_2)-\psi_n(x_2) \psi_{n'}(x_1)|^2\;,
  \label{eq:pdf}
\end{align}
where $\mathcal{N}_D= N(N-1)$. Since the terms with $n=n'$ in
eq.~\eqref{eq:pdf} vanish, and we can rewrite it in the following form
\begin{equation}
  \label{eq:pdf2}
    D(x_1,x_2)=\rho(x_1)\rho(x_2)-|\Delta(x_1,x_2)|^2\;,
\end{equation}
which is dependent only on the single particle density and the
correlation function $|\Delta(x_1,x_2)|$ defined by
\begin{equation}
  \label{eq:corr}
    |\Delta(x_1,x_2)|=\sum_{0\leq n\leq n'\leq N-1}^{N-1} 
                      \psi_n^\ast(x_1) \psi_n(x_2)\;.
\end{equation}
The pair distribution functions for samples consisting of $N=5,10$ and
$30$ particles and for barrier heights of $\kappa=0,1$ and $10$ are
shown in Fig.~\ref{fig:pdf}. The first striking feature inherent to
all situations is the absence of any probability along the diagonal,
which is due to the impenetrable nature of the atoms. As the splitting
strength is increased one observes the absence of probability for a
joint measurement along the cross defined by the lines $x_1=0$ and
$x_2=0$, which is again due to the central position of the
$\delta$-splitting.

\section{Ground state occupation numbers and coherence effects}
\label{sect:occu}
The fraction of particles that are in the $\phi_0(x)$ orbital $f$ is
related to the largest eigenvalue $\lambda_0$ of the RSPDM by
$f=\frac{\lambda_0}{N}$. Therefore, in analogy to the macroscopic
occupation of a single eigenstate in a Bose-Einstein condensate, this
orbital is sometimes referred to as the {\sl BEC} state and the
quantity $\lambda_0$ hence acts as a measure of the coherence in the
system.  Recently Forrester {\sl et al.} have shown that, as one
increases the particle number, $\lambda_0/N$ tends toward $1/\sqrt{N}$
in the harmonically trapped case \cite{Forrester:03}. Here we will
show how the introduction of a central barrier affects this
$1/\sqrt{N}$ behavior in a dramatic way.

\subsection{Occupation numbers }
\label{subsect:occfluc}

\begin{figure}[t]
   \includegraphics[width=\linewidth]{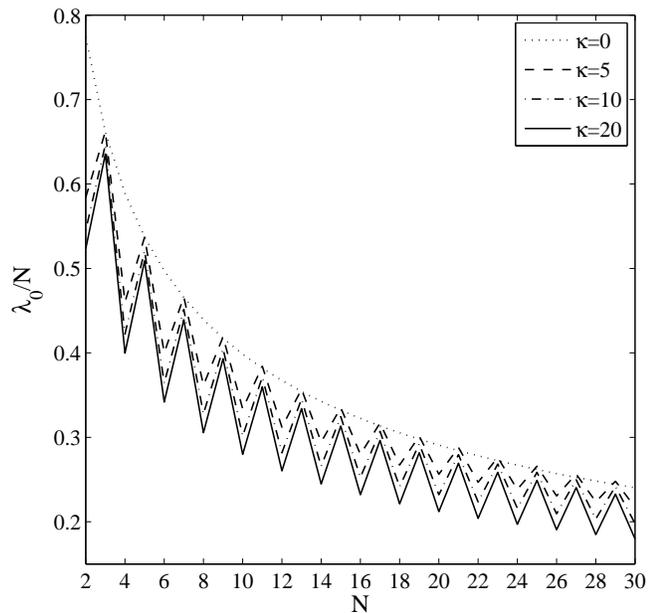}
   \caption{\label{fig:occ} Plot of the fractional ground state
     occupation $f=\frac{\lambda_0}{N}$ versus particle number for a
     TG gas in a $\delta$-split trap with increasing splitting
     strength.}
\end{figure}

The fractional ground state occupation $\lambda_0/N$ is displayed in
Fig.~\ref{fig:occ} as a function of particle number for different
heights of the splitting potential. The dotted line corresponds to the
unsplit trap ($\kappa=0$) and agrees with previously published results
\cite{Triscari:01,Forrester:03}. When increasing the magnitude of the
splitting one notes a strikingly different behavior for the values of
$\lambda_0/N$ for odd and even particle numbers. For even particle
numbers we find the coherence decreased as compared to the $\kappa=0$
case, whereas it is essentially unchanged when the particle number is
odd. The effect becomes more pronounced as the splitting strength is
increased and it damps out as the particle number is increased. This
behaviour is directly related to the magnitude of the RSPDM in the
off-diagonal quadrants, $\text{sgn}(x)\neq\text{sgn}(x')$ and
therefore a clear signature of the coherence inherent in the
system. To see this behaviour manifest itself in an experimentally
realizable quantity we will next calculate the momentum distribution
as well as the visibility of interference fringes in the free evolution
of the gas when released from the trap.

\subsection{Momentum distributions}
\label{subsect:RSPD}

For a harmonic trap the relationship between the momentum distribution
and coherence properties of the TG was recently studied by Minguzzi
and Gangardt \cite{Minguzzi:05}. The momentum distribution,
$n(k)$, can be calculated from the reduced single-particle density
matrix
\begin{equation}
 \label{eq:mom_dist_int}
 n(k) \equiv (2\pi)^{-1} \int_{- \infty}^{+ \infty}\int_{- \infty}^{+ \infty} 
 \rho (x,x')e^{- \imath k(x-x')} dx\; dx'\;,
\end{equation}
and is normalised as $\, \int_{- \infty}^{+ \infty} n (k) dk = N \,$.
Equivalently it can be obtain by considering the eigenstates of
$\rho(x,x')$.  Using a discretised form for the quadrature then allows
one to rewrite the integral equation as a linear algebraic equation
\begin{equation}
 \label{eq:mom_dist}
 n(k) = \sum_i\lambda_i| \mu_i(k) |^2\;,
\end{equation}
where $\mu_i(k)$ denotes the Fourier transform of the natural orbital
$\phi_i(x)$,
\begin{equation}
  \label{eq:ft_nat_orb}
  \mu_i(k)=\frac{1}{\sqrt{2 \pi}}\int_{-\infty}^{+\infty} 
  \phi_i(x)e^{- \imath k x}dx\;.        
\end{equation}

\begin{figure}[t]
   \includegraphics[width=\linewidth]{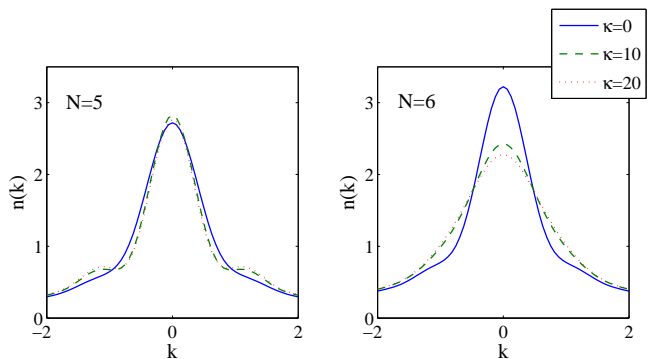}
   \caption{\label{fig:momplot} (Color online) Central peaks of the
     momentum distributions $n(k)$ for TG gases consisting of $N=5$
     (left) and $N=6$ (right) particles, for values of splitting
     strength $\kappa=0,10$ and $20$.}
\end{figure}

\begin{figure}[t]
   \includegraphics[width=\linewidth]{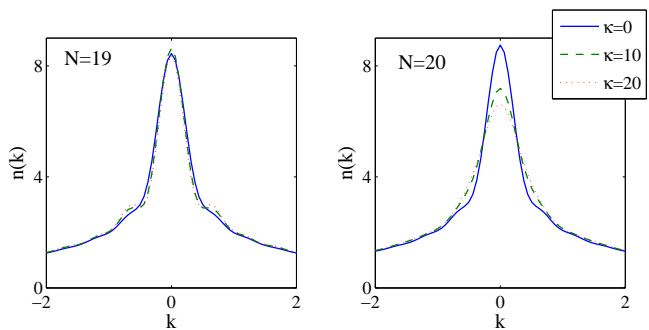}
   \caption{\label{fig:closemom} (Color online) Central peak of the 
momentum distribution $n(k)$ for TG gases consisting of $N=19$
     (left) and $N=20$ (right) particles, for values of splitting
     strength $\kappa=0,10$ and $20$.}
\end{figure}

\begin{figure}[t]
   \includegraphics[width=\linewidth]{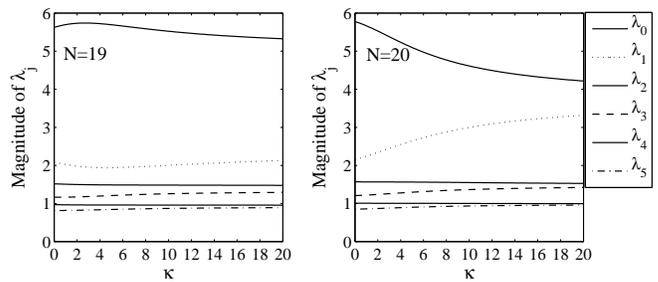}
   \caption{\label{fig:lam} Magnitude of the first six eigenvalues $\lambda_j$ of $\rho(x,x')$ for $N=19$ and $N=20$ particles as a function of the splitting strength $\kappa$. The lines of greatest magnitude represent $\lambda_0$ and decrease in the order $\lambda_0,\lambda_1,\lambda_2,\lambda_3,\lambda_4,\lambda_5$. }
\end{figure}

The central peaks of the reciprocal momentum distribution for $N=5$
and $N=6$ particle samples are shown in Fig.~\ref{fig:momplot} for
different values of the splitting strengths. Fig.~\ref{fig:closemom}
shows the same quantity for gases with $N= 19$ and $N=20$
particles. While the states with odd particle numbers are clearly less
affected by the barrier than the states with even particle number, one
can see the emergence of bi-modality at the neck of the peaks in the
plots on the left hand side. It stems from the interference of
particles on both sides of the splitting potential. For even particle
number the introduction of the central splitting significantly
broadens the momentum distribution lowers it's peak. This is in
agreement with analytical results we have found earlier for the
special case of a two-particle Tonks molecule \cite{Murphy:07} and
indicates a loss of coherence within the sample. The form of
the momentum distributions displayed in Fig.~\ref{fig:momplot} and Fig.~\ref{fig:closemom} is determined by
Eq.~\ref{eq:mom_dist}. In this equation $\lambda_j$ acts as a weight on the contribution of the Fourier transform of an individual natural orbital. It is interesting to see which natural orbitals are responsible for the broadening of the distribution in the even particle case. In Fig.~\ref{fig:lam} the first six $\lambda_j$'s are plotted as a function of $\kappa$ for $N=19$ and $N=20$ particles. The difference between the odd and even samples can be clearly seen. In the $N=20$ case we see a degeneracy of consecutive eigenvalues occuring, this is due to the symmetry of having ten particles on the right of the barrier and ten on the left. The degeneracy explains why we have a noticable change in the respective momentum distribution. In the $N=19$ case no such degeneracy occurs due to fact that one of the particles is spatially delocalised over both traps, this also explains the emergence of  bi-modality in the neck of the odd distributions. This behavior was
found to be consistent for all particle numbers studied, up to $30$ particles.

\subsection{Interference fringes}
\label{subsect:intfr}

\begin{figure}[t]
   \includegraphics[width=\linewidth]{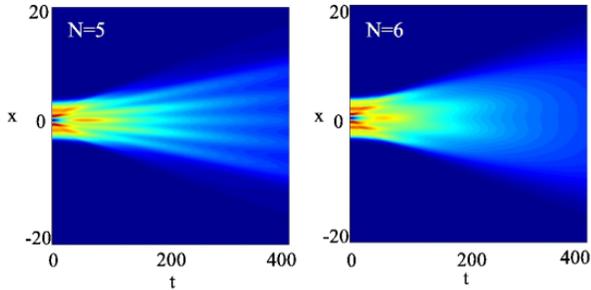}
   \caption{\label{fig:interference6j} (Color Online) Free temporal evolution of the
     single particle density, $\rho(x,t)$ for a $N=5$ and $N=6$
     particle TG gas initially confined in a $\delta$-split trap with
     $\kappa=100$.}
\end{figure}

\begin{figure}[t]
   \includegraphics[width=\linewidth]{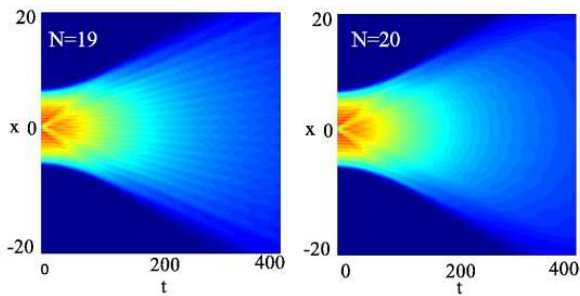}
   \caption{\label{fig:interference20j} (Color Online) Free temporal evolution of the
     single particle density $\rho(x,t)$ for a $N=19$ and $N=20$
     particle TG gas initially confined in a $\delta$-split trap with
     $\kappa=100$.}
\end{figure}

The Fermi-Bose mapping theorem holds equally well for time dependent
ground-state wave-functions and we will study the time evolution of the
many-body quantum state after removal of the external potential. This
situations is similar to one studied by Girardeau and Wright
\cite{Wright:00}, who considered splitting and recombining a gas
within an external trap. In our system splitting the sample is
inherent in the Hamiltonian and we will pay special attention to
effects stemming from different particle numbers. Starting off with
the gas confined in the $\delta$-split trap with large splitting
amplitude, we look at the time evolution of the single particle
density $\rho(x,t)$ as both, the trap and the central splitting, are
turned off and the gas undergoes free temporal evolution. During this
both halves of the trap start overlapping and the densities for
samples with $N=5$ and $N=6$ particles and with $N=19$ and $N=20$
particles are shown in Figs.~\ref{fig:interference6j} and
\ref{fig:interference20j}, respectively. In both cases one can clearly
see a more distinct interference pattern emerging for odd particle
than for even particle samples, indicating that odd states carry
larger coherence. In agreement with Fig.~\ref{fig:occ} the coherence
effect is less pronounced at larger particle numbers but still clearly
visible as can be seen from Fig.~\ref{fig:interference20j}.

It is well worth to point out that as Figs.~\ref{fig:interference6j}
and \ref{fig:interference20j} show only density distributions, the
results apply equally to a gas of spin polarized fermions. In this
case, however, the interpretation of the variations of coherence with
respect to particle number is straightforward, as the splitting
introduces non-smooth kinks in the single-particle wave-function which
create the Slater determinant.

\section{ Conclusions}
\label{sect:conclusions}

Tonks gas has over the previous years shown to be an exciting and rich
system to study new physics in a controlled way due to its analytic
accessibility. Identifying and describing new potentials that take
advantage of this is therefore of large importance. In the present
work we have undertaken a thorough investigation of the many-body
properties of the Tonks-Girardeau gas in a $\delta$-split trap. We
have calculated the RSPDM as well as the pair distribution function
for various different splitting magnitudes and particle numbers and
identified the basic physical behaviour shown by the system.

From the RSPDMs we were able to study coherence properties of the gas
by determining the behaviour of the ground state eigenvalue,
$\lambda_0$, as a function of particle number. Our results show that
odd and even particle number samples obey different scaling laws, with
the odd number samples remaining more coherent or less sensitive to
the central splitting. The effect becomes less pronounced as one goes
to larger particle numbers. To show how this effect manifest itself in
different and experimentally observable quantities, we have studied
the momentum distribution and interference experiments. For the
momentum distributions we found that for odd particle numbers the
sharp peak around momentum zero is relatively insensitive to the
different strength of the splitting. For even particle numbers,
however, the distributions are lowered and widened with increasing
$\kappa$, demonstrating a loss in coherence. The simulations of the
interference experiments for odd and even samples showed a larger
visibility occurring for odd particle samples. This is in agreement with
the other quantities that the odd number samples are more phase
coherent.

\begin{acknowledgements}
  JG would like to thank D.~O' Donoghue for valuable discussions. This
  project was supported by Science Foundation Ireland under project
  number 05/IN/I852.
\end{acknowledgements}

\end{document}